\documentclass{iaus}
\usepackage{graphicx}

\title[Photoionization modeling of Galactic planetary nebulae] 
{Photoionization modeling of the Galactic planetary nebulae Abell\,39
and NGC\,7027}

\author[A.~Danehkar, D.\,J.~Frew, Q.\,A.~Parker \& O.~De\,Marco]   
{A.~Danehkar$^1$, D.\,J.~Frew$^1$, Q.\,A.~Parker$^{1,2}$ \and
O.~De\,Marco$^1$}

\affiliation{$^1$Department of Physics and Astronomy, Macquarie University, Sydney, NSW 2109, Australia \\[\affilskip]
$^2$Australian Astronomical Observatory, PO Box 296, Epping, NSW
1710, Australia\\email: {\tt ashkbiz.danehkar@mq.edu.au;
david.frew@mq.edu.au}}

\pubyear{2011}
\volume{283}  
\pagerange{119--126}
 \jname{Planetary Nebulae: an Eye to the
Future} \editors{A.~Manchado, L.~Stanghellini \& D.~Schoenberner,
eds.}
\begin{document}

\maketitle

\begin{abstract}
We estimate distances to the spherical planetary nebula Abell\,39
and the bipolar planetary nebula NGC\,7027 by interpolating from a
wide grid of photoionization models using the 3-D code, MOCASSIN. We
find preliminary distances of 1.5\,kpc and 0.9\,kpc respectively, with uncertainties of about 30\%.

\end{abstract}

\firstsection 
\section{Introduction}

Accurate distances to planetary nebulae (PNe) are crucial in
unraveling the connection between the physical properties of the
nebulae with those of their central stars (CSs). Reliable distances
facilitate the accurate estimation of  fundamental parameters such
as the CS mass and luminosity, and the nebular mass and age.  We
have begun a program of using photoionization modeling to refine the
distances to a sample of nearby PNe (see also \cite[Danehkar \etal\
2011]{Danehkar_etal11}).
In this work, we study two very different PNe as a proof of concept:
Abell\,39, a simple spherical shell with no microstructures, and
NGC\,7027, a well-known, young, luminous bipolar PN with a massive
molecular envelope. Using the 3-D photoionization code (MOCASSIN;
\cite[Ercolano et al. 2003]{Ercolano_etal03}),
our ultimate aim is to constrain the distance to individual PNe, utilizing the physical PN
radius we have calculated and the angular size.

\section{Analysis}

Our procedure allows us refine the nebular parameters
iteratively to provide the best match against the observations, though owing to degeneracies in these parameters,
there may be more than one unique solution.  Since a
black body is not a perfect model for the CS continuum flux, we used
various NLTE model atmosphere
fluxes from the grid of \cite[Rauch (2003)]{Rauch_03}.
For the initial model inputs for A~39, we adopt the line
intensities and abundances from \cite[Jacoby \etal\
(2001)]{Jacoby_etal01}, and derived the nebular $T_{\rm e}$
from the [O\textsc{iii}] lines
Because the density of A~39 is very low, we initially determined
$N_e$ from the PN diameter and integrated H$\beta$ flux. Our
preliminary distance estimate is $D$ = 1.5~kpc, and the CS
parameters are $T_{\rm eff} = 160$\,kK and $L/L_{\odot} = 1800$,
which disagrees with the estimate of $T_{\rm eff}$ = 117\,kK from
Ziegler (2011, these proceedings).  The higher temperature is needed
to explain the observed He\,\textsc{ii} and [Ne\,\textsc{v}] line
intensities in the nebular shell.

\begin{figure}[tb]
\begin{center}
\includegraphics[width=5.1in ]
{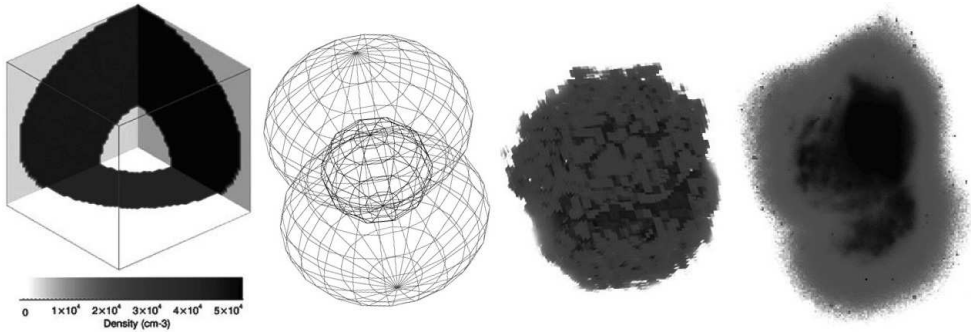}
\end{center}
\caption{(Left) Density distribution, cross section, and morphology
used for NGC\,7027. (Right) Computed surface brightness in the He
\textsc{ii} $\lambda$4686 line compared with the \textit{HST}
image.} \label{fig_1}
\end{figure}

For NGC\,7027, a starting value of $T_{\rm e}$ = 11\,kK is adopted,
the line intensities and abundances being taken from \cite[Bernard
Salas \etal\ (2001)]{BernardSalas_etal01} and \cite[Zhang \etal\
(2005)]{Zhang_etal05}, and refined as necessary. In addition, a
bipolar morphology with a mean density ($n_{\rm H} = 55,000\,{\rm
cm}^{-3}$; Figure \ref{fig_1}) is needed to model NGC\,7027, since a
simple assumption of spherical geometry conflicts with the
observed ionization structure and nebular line intensities.  
Our photoionization model outputs generally agree with the
observations, except for some uncertain lines.
Our photoionization model of NGC\,7027 gives $D$ = 900\,pc, with the
CSPN having $T_{\rm eff} = 180$\,kK and $L/L_{\odot} = 7500$, which
reproduces most observations. Our distance  agrees with previous
determinations (\cite[Masson 1989]{Masson89}; \cite[Volk \& Kwok
1997]{VolkandKwok97}; \cite[Zijlstra \etal\ 2008]{Zijlstra_etal08}).

\begin{table}[tbp]
\caption{Best-fit parameters (left) and observations versus model
outputs (right).} \label{table_lines}
\begin{center}
{\scriptsize
\begin{tabular}{l|c|c}
\hline \multicolumn{1}{l|} { Parameter } &  {\,A\,39} & {\,NGC\,7027}  \\
\hline
$T_{\rm eff}$ $({\rm K})$ & 160\,000 & 180\,000\\
$L_{\rm cs}$ $({\rm L_{\odot}})$& 1800 &  7500 \\
$M_{\ast}$ $({\rm M_{\odot}})$& 0.59 & 0.69\\
${R}_{\rm out} ({\rm pc})$& $0.64$ & 0.022 \\
$\delta{r}_{\rm shell} ({\rm pc})$& $0.07$ & 0.015\\
$D ({\rm pc})$& $1500$ &  900\\
$n_{\rm H}$ $({\rm cm}^{-3})$& $36$ &  $55\,000$ \\
$T_{e}$ $({\rm K})$& 15\,400  &  16\,400 \\
$\varepsilon$ & 0.45  &  0.50\\
${\rm log}({\rm He}/{\rm H})$& $-1.02$ & $-1.0$  \\
${\rm log}({\rm N}/{\rm H})$& $-4.30$ &  $-3.89$ \\
${\rm log}({\rm O}/{\rm H})$& $-3.76$  &  $-3.52$ \\
${\rm log}({\rm Ne}/{\rm H})$& $-4.45$ &  $-4.15$  \\
${\rm log}({\rm S}/{\rm H})$& $-5.19$ &  $-5.08$ \\
${\rm log}({\rm Ar}/{\rm H})$& $-6.05$ & $-5.80$ \\
\hline
\end{tabular}
}~~~{\scriptsize
\begin{tabular}{lc|cc|cc}
\hline \multicolumn{2}{c|} {Line } & \multicolumn{2}{c|} {A\,39} & \multicolumn{2}{c} {NGC\,7027} \\
Ion & $\lambda$({\AA}) & Obs. & Mod. & Obs. & Mod. \\
\hline
$[$Ne \textsc{v}$]$ & 3426
& 9.8 & 8.0 & 154 & 139 \\
$[$O \textsc{ii}$]$ & 3727
& 40 & 51 & 21  & 21\\
$[$Ne \textsc{iii}$]$ & 3869
& 104 & 92 & 126 & 143 \\
H$\gamma$ & 4340
& 48 & 47 & 47 & 48 \\
$[$O \textsc{iii}$]$& 4363
& 24 & 20 & 25 & 39 \\
He \textsc{ii}& 4686
& 95 & 94 & 49 & 62\\
$[$Ar \textsc{iv}$]$& 4740
& 4.0 & 4.5 & 8.1 & 8.1\\
H$\beta$ & 4861
& 100 & 100 & 100  &  100\\
$[$O \textsc{iii}$]$& 5007
& 1131 & 957 &  1397 & 1345\\
$[$N \textsc{ii}$]$& 5755
& ... & 1.4 & 5.8  & 7.2 \\
He \textsc{i}& 5876
& 1.9 & 1.9 &  10.9 & 7.5\\
H$\alpha$ & 6563
& 286 & 278 &  285 & 275 \\
$[$N \textsc{ii}$]$& 6584
& 12 & 41 &  110 &  118 \\
$[$S \textsc{ii}$]$& 6724
& 9.7 & 11 & 5.6 & 5.3  \\
$[$Ar \textsc{iii}$]$& 7135
& 5.8 & 2.2 & 21 & 8\\
\hline ${L}_{\rm H\beta}$(erg/s) &$1$E$33$
& 0.46 & 0.50 & 131 & 128 \\
\hline
\end{tabular}
}
\end{center}
\end{table}

\section*{Acknowledgments}
AD acknowledges receipt of an MQRES PhD Scholarship and an IAU Travel Grant.

\end{document}